\documentclass[amsmath,amssymb,amsfonts,aps,pre,preprint,superscriptaddress,bibnotes,showpacs,showkeys,longbibliography]{revtex4-1}

\usepackage{graphicx}
\usepackage{subfigure}
\usepackage{amsmath}
\usepackage{amsfonts}
\usepackage{amssymb}
\usepackage{natbib}
\usepackage{relsize}
\usepackage{sidecap}
\usepackage{braket}
\usepackage{footmisc}
\usepackage{footnote}
\usepackage{color}
\begin{document}

\title{Casimir force of a dilute Bose gas confined by a parallel plate geometry in improved Hatree-Fock approximation}

\author{Nguyen Van Thu and Pham The Song}
\affiliation{Department of Physics, Hanoi Pedagogical University 2, Hanoi, Vietnam}
\affiliation{Tay Bac University, Son La, Vietnam}

\begin{abstract}
Within framework of quantum field theory, in improved Hatree-Fock (IHF) approximation, we have considered a dilute single Bose-Einstein condensate (BEC) confined between two parallel plates. We found that the effective mass and order parameter of BEC strongly depend on distance separating two plates. Our results shows that the effective mass, order parameter and the Casimir force in IHF approximation equal to their values in one-loop approximation added a corrected term due to contribution of two-loop diagrams. We also show that the one-loop approximation is enough for calculating Casimir effect in an ideal Bose gas.
\end{abstract}

\maketitle

\section{Introduction\label{sec:1}}

T\label{sec:1}
In spite of discovering since 1948 \cite{Casimir}, the Casimir effect is and will be a cutting edge subject in modern physics. At first, it was discovered as a manifestation of the zero-point energy of an electromagnetic field. More than twenty years later, this effect has become more popular and attracted the attention of many physicists. The first noticeable work is Ref. \cite{Kay}, in which the Casimir effect in quantum theory was considered, especially for $\lambda\varphi^4$-theory. Up to now this effect has been investigated in various field of physics, such as, superconductor \cite{Wilson}, atomic and molecular physics \cite{Babb}, quark matter \cite{ThuPhat}, gravitation and cosmology \cite{Quach,Quach1} at both zero and non-zero temperature.

The Casimir effect in a Bose-Einstein condensate (BEC) the consequence of quantum fluctuations on top of ground state associating with phononic excitations \cite{Biswas,Biswas2,Biswas3,Schiefele}. At first it was considered for a single ideal BEC and shown that the Casimir force is attractive and proportional to $1/\ell^4$ at zero temperature, whereas at high temperature, it depends on both the distance $\ell$ between two parallel plates and temperature $T$  in form $T/\ell^3$ \cite{Biswas}. For the interacting Bose gas, in Ref. \cite{Schiefele} the Casimir force is expressed via integral of density of state. Instead of using Abel-Plana formula as in \cite{Schiefele}, Roberts and Pomeau \cite{Pomeau} employed Euler-MacLaurin formula and pointed out that at zero temperature, the Casimir force in a dilute interacting Bose gas depends on the coupling constant and decays as power law of the distance, which is well known result
\begin{eqnarray*}
F_C=-\frac{\pi^2}{480}\frac{\hbar v_s}{\ell^4},
\end{eqnarray*}
in which $\hbar$ is Planck's constant. The speed of sound $v_s$ depends on the interaction of atoms in the system. By analogous way, we obtained a relation for Casimir force for both Dirichlet and Robin boundary conditions \cite{Thu382}. For two-component Bose-Einstein condensates, Ref. \cite{Thu1} proved that  the Casimir force is not simple superposition of the one of two single component BEC due to the interaction between two species and one of the most important result  is that this force is vanishing in limit of strong segregation. However, the common feature of these papers is that the quantum field theory was used at one-loop approximation. In order to improve the results, at the same time, check the validity of these results, in this paper we keep up to double bubble of the quantum field theory to consider the Casimir effect in BEC. Moreover, one also employs a method proposed by \cite{Ivanov}, in which number of Goldstone bosons are preserved. This method is called improved Hatree-Fock approximation (IHF).

To begin with, we consider a dilute Bose gas described by Lagrangian \cite{Pethick},
\begin{eqnarray}
{\cal L}=\psi^*\left(-i\hbar\frac{\partial}{\partial t}-\frac{\hbar^2}{2m}\nabla^2\right)\psi-\mu\left|\psi\right|^2+\frac{g}{2}\left|\psi\right|^4,\label{1}
\end{eqnarray}
with $\psi=\psi(\vec{r},t)$ being the field operator, its average value plays the role of order parameter; $m$ is the atomic mass; the strength of repulsive intraspecies interaction is determined by coupling constant $g=4\pi\hbar^2a_s/m>0$ with $a_s$ being the $s$-wave scattering length \cite{Pitaevskii}. The system under consideration is connected to a bulk reservoir of condensate  hence the chemical potential has a constant value $\mu=gn_0$, where $n_0$ is bulk density of the condensate.

This paper is organized as follow. In Section \ref{sec:2} we present shortly about gap and Schwinger-Dyson equations for a single Bose gas in IHF approximation. The Casimir force is investigated in IHF approximation in Section \ref{sec:3}. Conclusion and Outlook given in Section \ref{sec:4} to close the paper.

\section{The effective potential in improved Hatree-Fook approximation}
\label{sec:2}
In this Section, we make a brief of process to obtain gap and Schwinger-Dyson (SD) equations for the BEC in IHF approximation. To do this, one first shifts the field operator as follow \cite{Phat},
\begin{eqnarray}
\psi\rightarrow \psi_0+\frac{1}{\sqrt{2}}(\psi_1+i\psi_2).\label{shift}
\end{eqnarray}
Plugging (\ref{shift}) into Lagrangian (\ref{1}) we get the interaction Lagrangian in double-bubble approximation
\begin{eqnarray}
{\cal L}_{int}=\frac{g}{2}\psi_0\psi_1(\psi_1^2+\psi_2^2)+\frac{g}{8}(\psi_1^2+\psi_2^2)^2.\label{Lint}
\end{eqnarray}
In tree approximation one has the gap equation
\begin{eqnarray}
\psi_0(-\mu+g\psi_0^2)=0,\label{gaptree}
\end{eqnarray}
and the inverse propagator
\begin{eqnarray}
D_0^{-1}(k)&=&\left(
              \begin{array}{cc}
                \frac{\hbar^2\vec{k}^2}{2m}+2g\psi_0^2 & -\omega_n \\
                \omega_n &  \frac{\hbar^2\vec{k}^2}{2m}\\
              \end{array}
            \right).\label{protree}
\end{eqnarray}
The Bogoliubov dispersion relation can be obtained by request that determinant of inverse propagator (\ref{protree}) is vanishing $\det D_0^{-1}(k)=0$ yielding
\begin{eqnarray}
E(k)=\sqrt{\frac{\hbar^2k^2}{2m}\left(\frac{\hbar^2k^2}{2m}+2g\psi_0^2\right)}.\label{dispertree}
\end{eqnarray}
It is obvious that this dispersion relation associates with Goldstone boson due to $U(1)\times U(1)$ breaking. We now write the CJT effective potential in Hatree-Fock approximation as pointed out in \cite{Phat} step by step
\begin{eqnarray}
V_\beta^{CJT}=&&-\mu\psi_0^2+\frac{g}{2}\psi_0^4+\frac{1}{2}\int_\beta \mbox{tr}\left[\ln D^{-1}(k)+D_0^{-1}(k)D(k)-{1\!\!1}\right]\nonumber\\
&&+\frac{3g}{8}(P_{11}^2+P_{22}^2)+\frac{g}{4}P_{11}P_{22},\label{VHF}
\end{eqnarray}
in which
\begin{eqnarray*}
P_{11}=\int_\beta D_{11}(k),~P_{22}=\int_\beta D_{22}(k),
\end{eqnarray*}
and
\begin{eqnarray*}
\int_\beta f(k)=T\sum_{n=-\infty}^{+\infty}\frac{d^3\vec{k}}{(2\pi)^3}f(\omega_n,\vec{k}).
\end{eqnarray*}
Here we denote $T$ for temperature and $\omega_n=2\pi nT$ is Matsubara frequency.

Without complexity, we can easily check that the Goldstone theorem fails in the Hatree-Fock approximation. To restore the Goldstone boson, one employs the method developed in \cite{Ivanov}. By this way, an extra term $\Delta V$ will be added in the CJT effective potential and hence
\begin{eqnarray}
V_\beta^{CJT}=&&-\mu\psi_0^2+\frac{g}{2}\psi_0^4+\frac{1}{2}\int_\beta \mbox{tr}\left[\ln D^{-1}(k)+D_0^{-1}(k)D(k)-{1\!\!1}\right]\nonumber\\
&&+\frac{g}{8}(P_{11}^2+P_{22}^2)+\frac{3g}{4}P_{11}P_{22}.\label{VIHF}
\end{eqnarray}
The approximation with restored Goldstone boson is called IHF. Minimizing the CJT effective potential with respect to order parameter $\phi_0$ and elements of propagator one arrives the gap equation
\begin{eqnarray}
-\mu+g\psi_0^2+\Sigma_2=0,\label{gap}
\end{eqnarray}
and SD equation
\begin{eqnarray}
M^2=-\mu+3g\psi_0^2+\Sigma_1,\label{SD}
\end{eqnarray}
respectively. In these equations we denote
\begin{eqnarray}
\Sigma_1&=& \frac{g}{2}P_{11}+\frac{3g}{2}P_{22},\nonumber\\
\Sigma_2&=&\frac{3g}{2}P_{11}+\frac{g}{2}P_{22}.\label{sig}
\end{eqnarray}
Combining Eqs. (\ref{VIHF})-(\ref{sig}) one has the inverse propagator
\begin{eqnarray}
D^{-1}=\left(
              \begin{array}{lr}
                \frac{\hbar^2k^2}{2m}+M^2 & -\omega_n \\
                \omega_n & \frac{\hbar^2k^2}{2m} \\
              \end{array}
            \right).\label{proIHF}
\end{eqnarray}
The dispersion relation is attained
\begin{eqnarray}
E(k)=\sqrt{\frac{\hbar^2k^2}{2m}\left(\frac{\hbar^2k^2}{2m}+M^2\right)}.\label{disperIHF}
\end{eqnarray}

The momentum integrals will be calculated by using formulae
\begin{eqnarray*}
\sum_{n=-\infty}^{+\infty}\frac{1}{\omega_n^2+E^2(k)}&=&\frac{1}{2TE(k)}\left[1+\frac{2}{e^{E(k)/k_BT}-1}\right],\nonumber\\
T\sum_{n=-\infty}^{n=+\infty}\ln\left[\omega_n^2+E^2(k)\right]&=&E(k)+2T\ln\left[1-e^{-E(k)/k_BT}\right],
\end{eqnarray*}
where $k_B$ and $T$ are Boltzmann constant and temperature, respectively. It is easy to check that at zero temperature these integrals has the form
\begin{eqnarray}
&&P_{11}=\frac{1}{2}\int\frac{d^3\vec{k}}{(2\pi)^3}\sqrt{\frac{\hbar^2k^2/2m}{\hbar^2k^2/2m+M^2}},~P_{22}=\frac{1}{2}\int\frac{d^3\vec{k}}{(2\pi)^3}\sqrt{\frac{\hbar^2k^2/2m+M^2}{\hbar^2k^2/2m}},\nonumber\\
&&\Omega_j\equiv\frac{1}{2}\int_\beta \mbox{tr}\ln D^{-1}(k)=\frac{1}{2}\int\frac{d^3\vec{k}}{(2\pi)^3}\sqrt{\frac{\hbar^2k^2}{2m}\left(\frac{\hbar^2k^2}{2m}+M^2\right)}.\label{tichphan}
\end{eqnarray}

We now simplify above equations by introducing dimensionless quantities: wave vector $\kappa=k\xi$, effective mass ${\cal M}=M/\sqrt{gn_0}$ with $\xi=\hbar/\sqrt{2mgn_0}$ being the healing length. Based on this, the momentum integrals in Eqs. (\ref{tichphan}) have the form
\begin{eqnarray}
&&P_{11}=\frac{1}{2\xi^3}\int\frac{d^3\kappa}{(2\pi)^3}\frac{\kappa}{\sqrt{\kappa^2+{\cal M}^2}},~P_{22}=\frac{1}{2\xi^3}\int\frac{d^3\kappa}{(2\pi)^3}\frac{\sqrt{\kappa^2+{\cal M}^2}}{\kappa}.\label{tichphan1a}
\end{eqnarray}
Introducing the dimensionless order parameter $\phi=\psi_0/\sqrt{n_0}$, Eqs. (\ref{gap}) and (\ref{SD}) can be rewritten as
\begin{eqnarray}
&&-1+\phi^2+\frac{1}{gn_0}\Sigma_2=0,\label{gap1}\\
&&{\cal M}^2=-1+3\phi^2+\frac{1}{gn_0}\Sigma_1.\label{SD1}
\end{eqnarray}
These equations allow us to study properties of the BEC.

\section{Casimir force in improved Hatree-Fock approximation}
\label{sec:3}

We now consider the case, in which Bose gas is confined between two parallel plates, which perpendiculars to $z$-axis at distance $\ell$. Along $0x,0y$ directions, our system is translational. The finite size caused by plates leads to quantization of the wave vector in $z$-component
\begin{eqnarray*}
k^2\rightarrow k_\perp^2+k_n^2,
\end{eqnarray*}
in which the wave vector component $k_\perp$ perpendiculars to $0z$-axis and $k_n$ is parallel with $0z$-axis. For boson system the periodic boundary condition is employed
\begin{eqnarray*}
k_n=\frac{2\pi n}{\ell},~n\in{\mathbb{Z}}.
\end{eqnarray*}
In dimensionless form one has
\begin{eqnarray}
\kappa^2\rightarrow \kappa_\perp^2+\kappa_j^2, ~\kappa_{n}=\frac{2\pi n}{L}\equiv\frac{n}{\overline{L}},~\overline{L}=\frac{L}{2\pi}.\label{k1}
\end{eqnarray}
Because of (\ref{k1}), the momentum integrals in Eq. (\ref{tichphan1a}) become
\begin{eqnarray}
P_{11}&=&\frac{1}{2\xi^3}\sum_{n=-\infty}^{+\infty}\int \frac{d^2\kappa_n}{(2\pi)^2}\sqrt{\frac{\kappa_\perp^2+\kappa_n^2}{\kappa_\perp^2+\kappa_n^2+{\cal M}^2}},\nonumber\\
P_{22}&=&\frac{1}{2\xi^3}\sum_{n=-\infty}^{+\infty}\int \frac{d^2\kappa_n}{(2\pi)^2}\sqrt{\frac{\kappa_\perp^2+\kappa_n^2+{\cal M}^2}{\kappa_\perp^2+\kappa_n^2}}.\label{k2}
\end{eqnarray}
By introducing a momentum cut-off $\Lambda$, the integrals in Eqs. (\ref{k2}) reduce to
\begin{eqnarray}
P_{11}&=&\frac{1}{4\pi\xi^3}\int_0^\Lambda \kappa_\perp d\kappa_\perp\sum_{n=-\infty}^{+\infty} \sqrt{\frac{\kappa_\perp^2+\kappa_n^2}{\kappa_\perp^2+\kappa_n^2+{\cal M}^2}},\nonumber\\
P_{22}&=&\frac{1}{4\pi\xi^3}\int_0^\Lambda \kappa_\perp d\kappa_\perp\sum_{n=-\infty}^{+\infty} \sqrt{\frac{\kappa_\perp^2+\kappa_n^2+{\cal M}^2}{\kappa_\perp^2+\kappa_n^2}}.\label{k3}
\end{eqnarray}
To deal with the sum in (\ref{k3}) the Euler-Maclaurin formula \cite{Arfken} is employed
\begin{eqnarray}
\sum_{n=0}^\infty \theta_nF(n)-\int_0^\infty F(n)dn=-\frac{1}{12}F'(0)+\frac{1}{720}F'''(0)-\frac{1}{30240}F^{(5)}(0)+\cdots,\label{EM}
\end{eqnarray}
with
\begin{eqnarray*}
\theta_n=\left\{
           \begin{array}{ll}
             1/2, & \hbox{if $n=0$;} \\
             1, & \hbox{if $n>0$,}
           \end{array}
         \right.
\end{eqnarray*}
then one takes the limit $\Lambda\rightarrow\infty$. Among the others, the momentum integrals (\ref{k2}) give
\begin{eqnarray}
P_{11}=-\frac{mgn_0\pi^2\xi^2}{90\hbar^2{\cal M}\ell^3}, ~P_{22}=\frac{mgn_0{\cal M}}{12\hbar^2\ell}-\frac{mgn_0\pi^2\xi^2}{90\hbar^2{\cal M}\ell^3}.\label{k4}
\end{eqnarray}
Substituting (\ref{k4}) into (\ref{gap}) and (\ref{SD}) we have the gap and SD equations in dimensionless form
\begin{eqnarray}
-1+\phi^2+\frac{mg{\cal M}}{24\xi\hbar^2L}-\frac{mg\pi^2}{45\xi\hbar^2{\cal M}L^3}&=&0,\nonumber\\
-1+3\phi^2+\frac{mg{\cal M}}{8\xi\hbar^2L}-\frac{mg\pi^2}{45\xi\hbar^2{\cal M}L^3}&=&{\cal M}^2.\label{eq1}
\end{eqnarray}
\begin{figure}
  \includegraphics{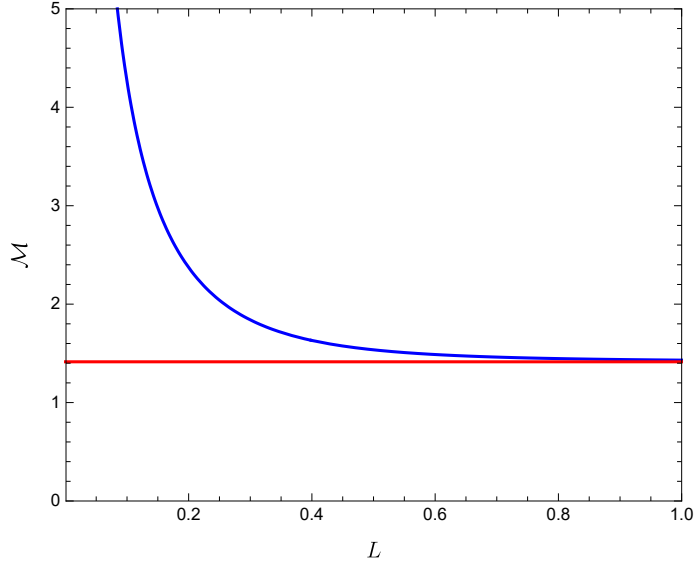}
\caption{(Color online) The dimensionless effective mass as a function of distance $L$. The red and blue lines correspond to one-loop and IHF approximation.}
\label{f1}       
\end{figure}
Solving these equations one obtains an analytical solution for the effective mass in dimensionless form
\begin{eqnarray}
{\cal M}=\sqrt{\frac{2}{3}}\left(\frac{1}{\widetilde{M}^{1/3}}+\widetilde{M}^{1/3}\right),\label{Mhd}
\end{eqnarray}
where
\begin{eqnarray*}
\widetilde{M}=\frac{4\pi^{7/2}n_s^{1/2}}{\sqrt{75}L^3}+\sqrt{\left(\frac{4\pi^{7/2}n_s^{1/2}}{\sqrt{75}L^3}\right)^2-1},
\end{eqnarray*}
in which $n_s=n_0a_s^3$. The first thing one can see is that the $\ell$-dependence of effective mass. This is a significant difference in comparing to that in one-loop approximation, where the effective mass is independent on the distance and have a constant value ${\cal M}_1=\sqrt{2}$ as in \cite{Pomeau,Thu1}. Fig. \ref{f1} shows $\ell$-dependence of the effective mass, the parameters are chosen associating rubidium Rb87 $m=86.909$ u, $a_s=500$ nm and $\xi=4000$ nm. It is obvious that the smaller distance $\ell$, the stronger $\ell$-dependence of effective mass and this effective mass is divergent when $\ell$ approaches to zero. Recall that we are considering the dilute Bose gas, i.e, $n_s=n_0a_s^3\ll1$
\cite{Andersen}, thus the effective mass can be expanded as
\begin{eqnarray}
{\cal M}\approx{\cal M}_1+\frac{4\sqrt{2}\pi ^{7/2} n_s^{1/2}}{45 L^3}.\label{Mhd1}
\end{eqnarray}
\begin{figure}
  \includegraphics{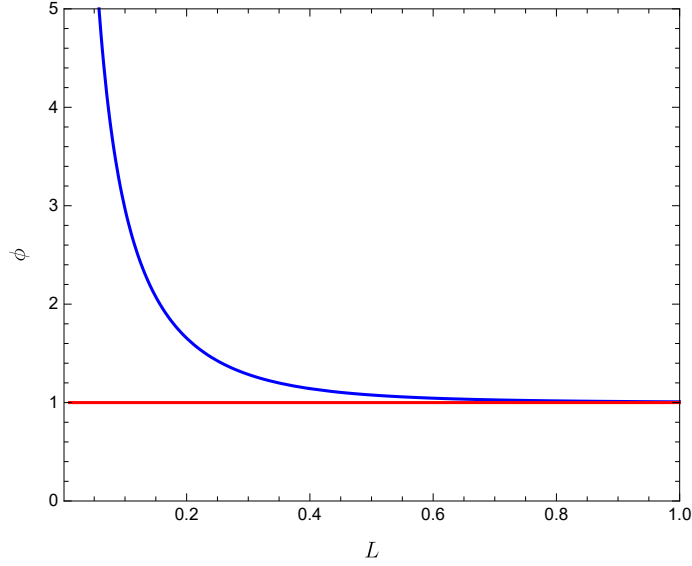}
\caption{(Color online) The evolution of order parameter versus distance. The red and blue lines correspond to one-loop and IHF approximation.}
\label{f2}       
\end{figure}

Based on the gap and SD equations (\ref{eq1}) we can also find the $\ell$-dependence of the order parameter and it is plotted in Fig. \ref{f2} with the same parameters in Fig. \ref{f1}, in which the blue curve corresponding to IHF approximation approaches to the red line associating on-loop approximation $\phi_1=1$ when the distance is large enough. The same as effective mass, the order parameter diverges at $L=0$. For a dilute gas one has
\begin{eqnarray}
\phi^2\approx\phi_1^2+2\pi^{3/2}\left(\frac{4\pi^2}{45L^3}-\frac{1}{3L}\right)n_s^{1/2}.\label{phi}
\end{eqnarray}

Let us now study on Casimir force due to the quantum fluctuations on top of ground state in IHF approximation at zero temperature. Within this framework the grand canonical energy density is
\begin{eqnarray}
\Omega=\frac{gn_{0}}{2\xi^3}\int\frac{d^3\kappa}{(2\pi)^3}\sqrt{\kappa^2(\kappa^2+{\cal M}^2)}.\label{energy}
\end{eqnarray}
Under the compactification of $z$-direction, the wave vector is quantized as (\ref{k1}) therefore the Casimir energy is finite part of the energy
\begin{eqnarray}
\Omega=\frac{gn_{0}}{2\xi^2\overline{L}^2}\sum_{n=-\infty}^\infty\int\frac{d^2\kappa_{\perp}}{(2\pi)^2}\sqrt{(\overline{L}^2\kappa_{\perp}^2+n^2)({\cal M}_0^2+n^2)},\label{term12}
\end{eqnarray}
in which
\begin{eqnarray*}
{\cal M}_0=\overline{L}\sqrt{\kappa_{\perp}^2+{\cal M}^2}.
\end{eqnarray*}
Using the Euler-Maclaurin formula (\ref{EM}) one has the Casimir energy \cite{Thu382},
\begin{eqnarray}
{\cal E}_C=-\frac{mg^2n_0^2\pi^2{\cal M}}{720\hbar^2L^3}.\label{energyC}
\end{eqnarray}
The Casimir force is attained by a derivative of Casimir energy with respect to distance between two plates
\begin{eqnarray}
F_C=-\frac{\partial {\cal E}_C}{\partial\ell}=-\frac{1}{\xi}\frac{\partial {\cal E}_C}{\partial L}.\label{dn}
\end{eqnarray}
\begin{figure}
  \includegraphics{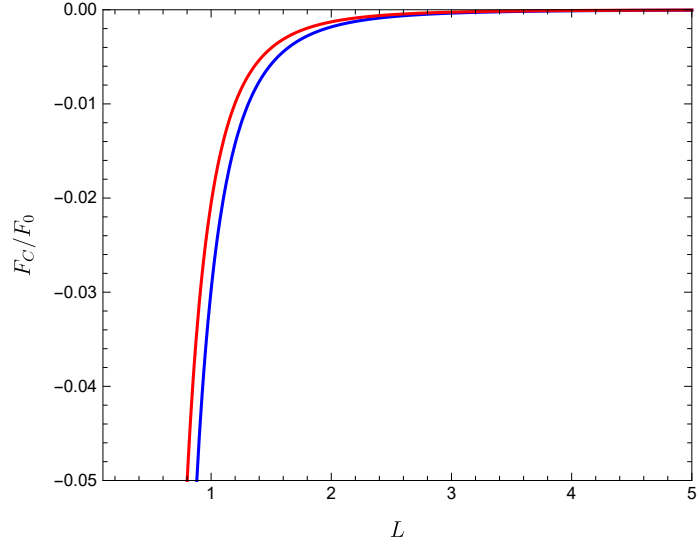}
\caption{(Color online) The Casimir force as a function of distance. The red and blue lines correspond to one-loop and IHF approximation.}
\label{f3}       
\end{figure}
Substituting (\ref{energyC}) and then into (\ref{dn}) we obtain
\begin{eqnarray}
F_C=F_0\frac{\pi^2}{1440}\frac{\partial}{\partial L}\left(\frac{{\cal M}}{L^3}\right),\label{dn1}
\end{eqnarray}
where $F_0=gn_0/\xi^3$. Combining (\ref{Mhd}) and (\ref{dn1}) one arrives
\begin{eqnarray}
\frac{F_C}{F_0}=-\frac{F_1}{720.\ 15^{2/3} L^7 \widetilde{M}^{1/3} \sqrt{\frac{32 \pi ^7n_s}{L^6}-150}},\label{force}
\end{eqnarray}
in which
\begin{eqnarray*}
F_1=\pi ^2 \left[3 L^3 \left(5.3^{1/3}+5^{1/3} \widetilde{M}^{2/3}\right) \sqrt{\frac{16 \pi ^7n_s}{L^6}-75}-4 \pi ^{7/2}n_s^{1/2} \left(5.3^{1/3}-5^{1/3}\widetilde{M}^{2/3}\right)\right].
\end{eqnarray*}
The evolution of Casimir force versus distance is shown in Fig. \ref{f3} for rubidium 87 with the same parameters as in Figs. \ref{f1} and \ref{f2}. It is clear that the Casimir force in IHF approximation is attractive and divergences when the distance approaches to zero. Like result of one-loop approximation, the Casimir force decays sharply and approaches to zero when distance $\ell$ is large enough. However, at a given value of the distance, strength of Casimir force in IHF approximation is always stronger that in one-loop approximation. To understand this, we write Eq. (\ref{dn1}) in form
\begin{eqnarray}
\frac{F_C}{F_0}=-\frac{\pi^2}{480L^4}+\frac{\pi^2}{1440L^3}\frac{\partial {\cal M}}{\partial L}\equiv {\cal F}_1+{\cal F}_{correct}.\label{dn2}
\end{eqnarray}
It is easily to recognize that the first term in right hand side of (\ref{dn2}) is the Casimir force in one-loop approximation \cite{Pomeau,Thu382} and second one is corrected term. For a dilute Bose gas, the corrected term can be written as
\begin{eqnarray}
{\cal F}_{correct}=-\frac{\pi^{11/2}n_s^{1/2}}{2700\sqrt{2}L^7}.\label{correct}
\end{eqnarray}
\begin{figure}
  \includegraphics{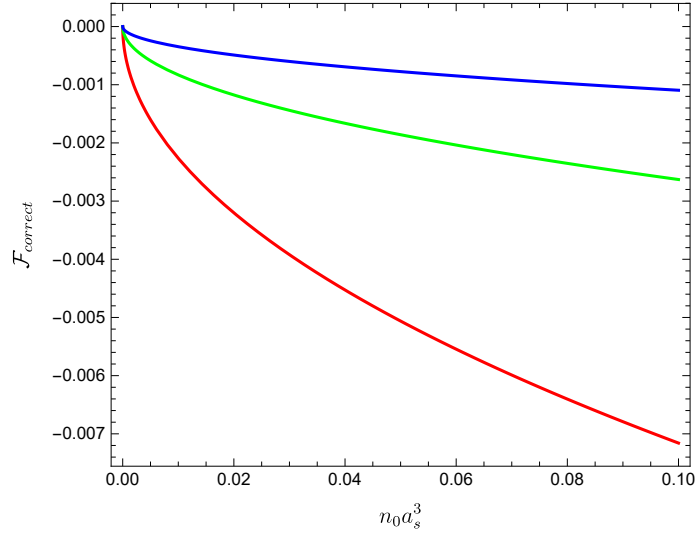}
\caption{(Color online) The corrected term versus $n_0a_s^3$ for sodium. The red, green and blue lines correspond to $L=1.3,1.5$ and 1.7, respectively.}
\label{f4}       
\end{figure}
Eq. (\ref{correct}) shows that the corrected term decreases very fast when the distance increases, thus its contribution is only appreciable at small-$\ell$. The evolution of the corrected term as a function of $n_s$ is shown in Fig. \ref{f4} for sodium with the same parameters as in Figs. \ref{f1}, \ref{f2} and \ref{f3}. The red, green and blue lines correspond to $L=1.3,1.5$ and 1.7, respectively. One also sees that at a given value of the distance $\ell$, the contribution of corrected term will increase if $n_s$ increases. At $n_s=0$ this contribution is vanishing. This means that, for a ideal Bose gas the one-loop approximation is enough to consider the Casimir force.

\section{Conclusion and Outlook}
\label{sec:4}

In this paper we investigate the theoretical calculations on effect of the finite size in one direction on the Casimir force of an interacting Bose gas within framework of IHF approximation. Our main results are in neat order

- We found analytical relations for the effective mass and order parameter. A vital difference in comparing to those in one-loop approximation is that these quantities strongly depend on the distance, especially in small distance region. By expanding these relations as power series of small quantity $n_s=n_0a_s^3$ (associating to a dilute Bose gas), we proved that in IHF approximation, the effective mass and order parameter equal to their values in one-loop approximation after adding a corrected terms.

- The Casimir force was also investigated in IHF approximation. The result shows that in small distance region the Casimir force diverges faster than that in one-loop approximation, the reason is the presence of corrected term, which stems from taking into account the higher order terms in interaction Lagrangian.

Last but not least, we confirmed that the one-loop approximation is enough for studying on Casimir force of an ideal Bose gas. Without interacting, the corrected term equals to zero for the ideal Bose gas.

Based on these results one can study on this problem on a Bose-Einstein-condensate mixtures to check some important results mentioned in Ref. \cite{Thu1}, especially the vanishing of Casimir force in limit of strong segregation. In addition, the net force consists of Casimir force and surface tension force \cite{Biswas3,Thu382} or Casimir-like force \cite{Thuphatsong} can also be considered in IHF approximation.

\section*{Acknowledgements}

 This work is funded by the Vietnam National Foundation for Science and Technology Development (NAFOSTED) under Grant
No. 103.01-2018.02. Authors are grateful to N. T. Lam for his stimulating discussion motivating Eq. (\ref{k4}). Prof. T. H. Phat and S. Biswas are acknowledged for their helpful remarks.

\section*{References}

\bibliography{mybibfile}

\end{document}